\documentclass[aps,apl,twocolumn,floatfig,10pt]{revtex4}

\usepackage{subfigure}
\usepackage{epsfig}
\usepackage{graphicx}
\usepackage{epstopdf}
\usepackage{subfigure}

\begin{document}
\title{An analytical model of Faraday rotation in hot alkali metal vapours}
\author{Stefan L Kemp, Ifan G Hughes and Simon L Cornish}
\affiliation{Department of Physics, University of Durham, South Road, Durham, DH1 3LE, United Kingdom}

\begin{abstract}
We report a thorough investigation into the absorptive and dispersive properties  of hot caesium vapour, culminating in the development of a simple analytical model for off-resonant Faraday rotation. The model, applicable to all hot alkali metal vapours, is seen to predict the rotation observed in caesium, at temperatures as high as 115~$^{\circ}$C, to within 1~\% accuracy for probe light detuned by greater than 2~GHz from the $D_{2}$ lines. We also demonstrate the existence of a weak probe intensity limit, below which the effect of hyperfine pumping is negligible. Following the identification of this regime we validate a more comprehensive model for the absorption and dispersion in the vicinity of the $D_{2}$ lines, implemented in the form of a computer code. We demonstrate the ability of this model to predict Doppler-broadened spectra to within 0.5~\% rms deviation for temperatures up to 50~$^{\circ}$C.
\end{abstract}

\maketitle
\section{Introduction}
The ability to control absorption and dispersion in hot alkali metal vapours is  of great interest in a number of areas of atomic and molecular physics. The field of slow-light, in particular, is very dependent on this ability \cite{EITreview,slowlight,slowlight2}. The control of these properties have other uses in, for example, light storage \cite{storage}, quantum memory \cite{qmemory}, entanglement of macroscopic systems \cite{contentanglement}, and frequency up-conversion \cite{Vernier}. In this work we will expand upon the flourishing interest in the Faraday effect. This phenomenon has already been put to many uses including: all-optical switching \cite{dawes, siddons4}, non-invasive atomic probing \cite{siddons2}, far off-resonance laser locking \cite{Marchant}, and realising a dichroic beam splitter \cite{Abel}.  A model for absorption and dispersion in alkali vapours will provide a tool for the prediction and analysis of many of these applications. It may also be used to predict the properties of EIT spectra \cite{eit2} and slow-light Fourier transform interferometers \cite{fourierint}.

The Faraday effect is a dispersive phenomenon manifesting itself in a frequency dependent rotation of the plane of linearly polarised light when an axial magnetic field is applied to an atomic vapour. This is due to the Zeeman shift of spectral lines driven by circularly polarised light. A significant advantage of using dispersive properties in atomic physics is that they dominate over absorption in off-resonant schemes \cite{siddons}. A model has been developed by Siddons \textit{et al.} for the absorption and dispersion in rubidium, which has since successfully predicted Doppler-broadened absorption \cite{siddons3} and dispersion using the Faraday effect \cite{siddons}. In addition to this Marchant \textit{et al.} have used the Faraday effect in order to lock lasers far from atomic resonance \cite{Marchant}. However none of these works characterise the significant temperature dependence of Faraday rotation; a property desired in many of its applications. 

The motivation of this study is to characterise the Faraday effect in caesium in off-resonant schemes. We adapt the model produced by Siddons \textit{et al.} in order to predict absorption and dispersion in caesium. This model is thoroughly tested through comparison to experimental Doppler-broadened absorption spectra. A result of this is the identification of a weak-probe regime in caesium, in which the effects of hyperfine pumping are negligible. Following the successful application of the model to the study of absorption, we are confident in the model's ability to predict dispersive effects. The model for the susceptibility involves Voigt profiles which require numerical evaluation using a computer code. However, the writing, testing an implementation of such a code is complicated and time-consuming. The aim of this work is to devise an analytic model for Faraday rotation in off-resonant schemes, greatly improving our understanding of the phenomenon and aiding applications such as laser locking far-detuned from resonance. This paper is structured as follows. In section two we present the theory behind the existing model. Section three provides details on the experimental methods used in the testing process. In section four we display our results, discuss their implications, and introduce a simple model for off-resonance Faraday rotation. Finally, in section five, we summarise and conclude.

\section{Absorption and Dispersion in Atomic Vapours}
The transmission of a weak monochromatic light source propagating through an atomic vapour is governed by the Beer-Lambert law; \begin{equation}I=I_{0}\exp \left[-\alpha\left(\nu,T\right)L\right],\end{equation} where \emph{I} is the transmitted probe intensity for a given incident intensity $I_{0}$. The absorption coefficient, $\alpha$, is dependent on both the frequency of the light, $\nu$, and the vapour temperature, \emph{T}. \emph{L} is the length of the cell containing the vapour. Here we assume that the light source is weak enough to not influence $\alpha$. The fraction of light transmitted, $\tau$, is then \begin{equation} \tau=\frac{I}{I_{0}}=\exp\left[-\alpha L\right].\end{equation}

In this paper we are primarily concerned with atom-light interactions at frequencies in the vicinity of the $D_{2}$ lines for caesium. These lines and the allowed electric dipole interactions among them are shown in figure \ref{setup}. Each of the many transitions in this frequency domain  contribute to $\alpha$, hence we may express the absorption coefficient as a sum \begin{equation}\alpha\left(\nu,T\right)=\sum_{j}\alpha_{j}\left(\nu,T\right),\end{equation} where $\alpha_{j}\left(\nu,T\right)$ is the absorption coefficient of transition \emph{j} at frequency $\nu$ and temperature \emph{T}.

In addition to $\alpha$, each transition also contributes to the refractive index, $\eta$, of the vapour. The $\sigma^{+}$ and $\sigma^{-}$ transitions in an atomic system are driven by left- and right-circularly polarised light respectively. When a magnetic field, \emph{B}, is applied along the propagation direction, the resonant frequencies of the $\sigma^{+}$ and $\sigma^{-}$ transitions are shifted in opposite directions by the Zeeman effect.  Consequently the refractive index profiles $\eta^{+}$ and $\eta^{-}$, associated with left- and right-circularly polarised light respectively, are also shifted in opposite directions. Since the linearly polarised light incident on the vapour can be considered as a superposition of the two circular polarisations, each of these components will therefore propagate at different speeds. Upon emerging from the cell, there will be a relative phase between the components of \begin{equation}\Delta\varphi=\frac{2\pi}{\lambda}L\left(\eta^{+}-\eta^{-}\right),\end{equation} where $\lambda$ is the wavelength of the probe beam. In the case of negligible absorption, the plane of polarisation rotates by \begin{equation}\theta=\frac{\pi}{\lambda}L\left(\eta^{+}-\eta^{-}\right),\label{angle}\end{equation} with respect to the plane of polarisation of the light incident on the cell.

The electric susceptibility, $\chi$, encapsulates both the absorption coefficient and the refractive index, and is a complex quantity; \begin{equation}\chi=\chi_{r}+i\chi_{i},\end{equation} where $\chi_{r}$ and $\chi_{i}$ are the real and imaginary parts of the susceptibility respectively. The absorption coefficient is characterised by \begin{equation}\alpha\left(\Delta\right)=k\chi_{i}\left(\Delta\right),\end{equation} where \emph{k} is the wavenumber and $\Delta$ is the detuning. Throughout this paper, unless specified otherwise, we treat zero detuning as the centre of mass frequency of the caesium $D_{2}$ lines. The refractive index is given by \begin{equation}\eta\left(\Delta\right)=\sqrt{\chi_{r}\left(\Delta\right)+1}.\end{equation} In a Doppler-broadened atomic vapour, $\chi$ for the transition $j$ is given as \begin{equation}\chi_{j}\left(\Delta\right)=c_{j}^{2}\frac{d^{2}N}{\hbar\epsilon_{0}}s_{j}\left(\Delta\right),\label{susceptibility}\end{equation} where $c_{j}^2$ is the transition strength and the \emph{N} is the atomic number density \cite{siddons3}. $s_{j}\left(\Delta\right)$ is the lineshape of the resonance; a convolution of the Lorentzian atomic lineshape and the Gaussian velocity distribution along the propagation direction. The total susceptibility is then calculated by summing over all transitions. In the first half of this work we extend the successful model of $\chi$ in rubidium developed by Siddons \textit{et al.} \cite{siddons3} by implementing the same computational methods for atomic caesium.

\section{Experimental Methods}
\subsection{Experimental set up}
\begin{figure}[ht]
\centering
\includegraphics*[width=\linewidth]{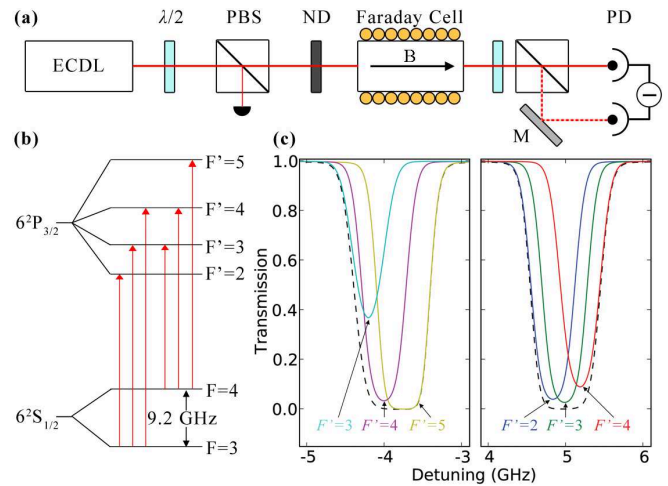}
\caption{(a) A schematic showing the experimental set up. Key: half wave plate ($\lambda$/2), polarising beam splitter (PBS), mirror (M), and differencing photodiode (PD). (b) The hyperfine energy states and transitions that make up the $D_{2}$ lines. (c) The theoretical $D_{2}$ transmission in a 7.5~cm caesium vapour
cell at 27~$^{\circ}$C (bottom right). The \emph{F}=4 line is shown on the left, and the \emph{F}=3 line on the right. The
contributions to the absorption of each hyperfine transition are shown and labelled. The
dashed black curve shows the total transmission through the cell.}
\label{setup}
\end{figure}

A schematic of the apparatus used is shown in figure \ref{setup}. An extended cavity diode laser at 852~nm was the light source used. The combination of a half-wave plate ($\lambda$/2), a polarisation beam splitting cube (PBS) and neutral density filters (ND) were used to create linearly polarised light of variable intensity. The probe beam then passed through a 7.5~cm heated caesium vapour cell based on the design of McCarron \textit{et al.} \cite{mccarron}. An external solenoid was used to provide both heating, and an axial magnetic field if required. Upon exiting the cell, the probe passed through a second half-wave plate and the two orthogonal linearly polarised components of the beam were separated using an analysing PBS cube. These components were then focussed onto the ports of a differencing photodiode.

The predictions of the computational code were validated by comparisons to the experimental Doppler-broadened transmission of the weak probe propagating through hot caesium vapour. These measurements were taken under zero magnetic field with the analysing half-wave plate set so that all transmitted light was incident on a single port of the differencing photodiode. Spectra were recorded for both $D_{2}$ lines at a range of temperatures using the heating solenoid.

Faraday signals were obtained by setting the analysing half-wave plate so that its optical axes were at $\pi$/4 to those of the analysing PBS cube. The output of the differencing photodiode then exhibited zero signal at detunings where the plane of polarisation was rotated by integer multiples of $\pi$/2.

\subsection{Scan Linearisation}
The ECDL used during this experiment incorporated current feed forward circuitry. This synchronised modulation of the diode current with the piezo allowed scan ranges of up to 12~GHz to be obtained. Although the laser frequency scans were driven by a triangle waveform, some linearisation was still required. The frequency axis of all scans recorded for this work were scaled by the use of a Fabry-Perot etalon and saturated absorption / hyperfine pumping spectroscopy \cite{macadam, smith}. The differences between the observed and expected detunings of the etalon transmission peaks were seen to follow a  low-order polynomial relation as a function of detuning. Subtracting a fit to the residuals from the measured detuning linearised the scans. The sub-Doppler references generated, along with crossover resonances at detunings halfway between each pair of hyperfine lines, were at well-known frequencies. These features were used to determine the absolute detunings across the scan.

\section{Results and Discussion}
\subsection{The Importance of a Weak Probe}
\begin{figure}[ht]
\centering
\includegraphics*[width=\linewidth]{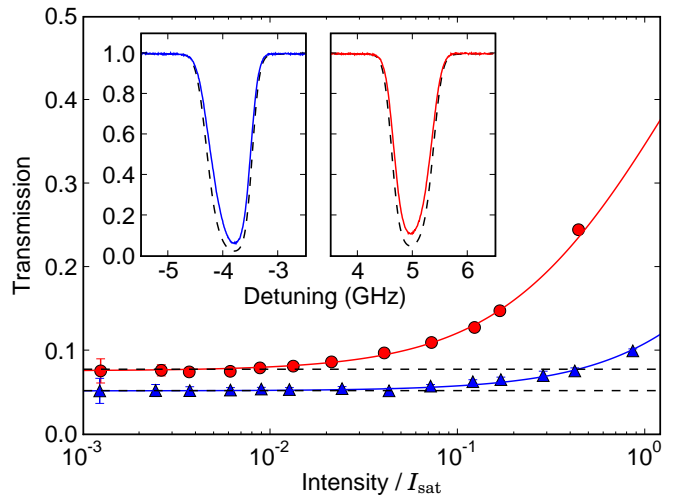}
\caption{Linecentre transmission as a function of probe intensity for the \emph{F}=3 (red circles) and \emph{F}=4 (blue triangles) lines. The corresponding curves are fits to equation \ref{weak}. The horizontal dashed black lines show the theoretical transmission in the weak probe regime for each line. The insets show the transmission across each line at 19~$^{\circ}$C with a probe intensity of 0.07~$I_{{\rm sat}}$. The dashed black curves show the theoretical transmission.}
\label{weakprobe}
\end{figure}

The inset to figure \ref{weakprobe} shows experimental absorption spectra of both the \emph{F}=3 and \emph{F}=4 lines at a temperature of 19~$^{\circ}$C. These spectra were taken with a probe intensity of 0.07~$I_{{\rm sat}}$, where the saturation intensity is 2.7~mW\thinspace cm$^{-2}$ \cite{corney}. It is clear to see that, at this intensity, the predictions of the code underestimate the levels of transmission observed in experiment. The \emph{F}=3 transition in particular shows very poor agreement across the detunings spanned by the line.

This occurs even though the probe intensity is below the transition's saturation intensity. An underlying assumption of the model is the cause of this; that the distribution of atoms in the hyperfine ground states is unchanged by the beam. In reality, it is possible for an atom excited from one hyperfine ground state, to decay spontaneously into the other ground state, which is known as a dark state since is is detuned by 9.2~GHz from the resonance being probed. This effect is called hyperfine pumping \cite{sherlock} and is the cause of the reduced absorption exhibited in figure \ref{weakprobe}.

The high frequency side of the \emph{F}=4 transition in the figure is clearly less strongly affected by this phenomenon. This is due to the closed transition to $F'$=5 dominating the lineshape, as may be seen in figure \ref{setup}. However the weaker closed transition to $F'$=2 in the \emph{F}=3 line does not have such a significant influence on the lineshape. Consequently hyperfine pumping has a strong impact over the entire line.

It has been seen in rubidium, that if the probe beam is sufficiently weak, then the time scale of hyperfine pumping becomes very long and the atoms may traverse the beam without being transferred to the dark state \cite{sherlock}. To investigate the effect of hyperfine pumping in the caesium $D_{2}$ lines, the linecentre transmission of each line was recorded across a large range of probe intensities. The results of this study are presented in figure \ref{weakprobe}. Included in this plot are guides to the eye, with the absorption coefficient modelled as \begin{equation}\alpha=\alpha_{{\rm weak}}\left(1+\beta\frac{I}{I_{{\rm sat}}}\right)^{-1/2},\label{weak}\end{equation} where $\alpha_{{\rm weak}}$ is the theoretical value of $\alpha$ in the absence of the pumping. $\beta$ is a fitting parameter characterising the change in the saturation intensity due to hyperfine pumping. It is immediately apparent from this plot that, at intensities below 0.01~$I_{{\rm sat}}$, the experimental transmission converges to the theoretical value. Despite this it is worth noting that this value of the weak-probe limit is valid only for the beam used in this experiment, since hyperfine pumping will depend upon both the beam size and power. However, this is still an important result, since we have demonstrated conclusive evidence of the existence of a weak-probe regime for the caesium $D_{2}$ lines.

\subsection{Comparison of Experiment to Theory}
\begin{figure}[ht]
\centering
\includegraphics*[width=\linewidth]{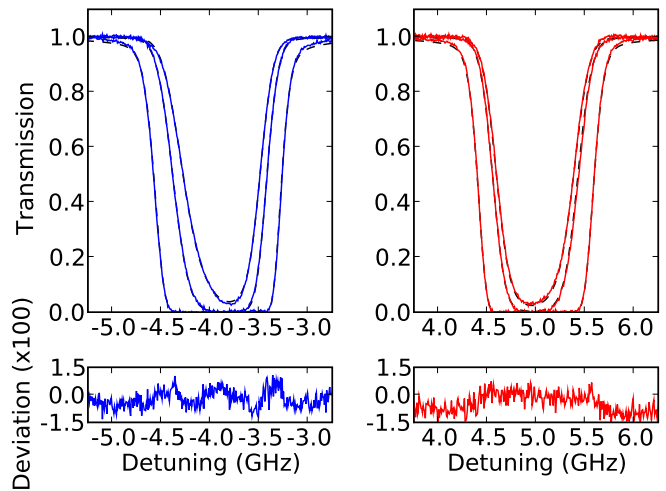}
\caption{The left plot shows transmission as a function of linear detuning for the \emph{F}=4 line at temperatures of 19~$^{\circ}$C (top curves), 25~$^{\circ}$C (middle) and 47~$^{\circ}$C (bottom). The right hand plot shows transmission for the \emph{F}=3 line at 21~$^{\circ}$C (top), 25~$^{\circ}$C (middle), and 47~$^{\circ}$C (bottom). The sub-plots show the experimental residuals from the model multiplied by a factor of 100 for the 25~$^{\circ}$C data.}
\label{results}
\end{figure}

A series of Doppler-broadened profiles of both $D_{2}$ lines have been recorded in order to test the model against experiment. These are shown in figure \ref{results} and were taken with a probe intensity of 0.001~$I_{{\rm sat}}$, which, from figure \ref{weakprobe}, is discernibly within the weak-probe limit. These spectra were recorded at a variety of different temperatures using the heating solenoid. The spectral lines clearly show the expected increase in absorption and width as the temperature is raised. In addition to the experimental data, the theoretical predictions of the model are also shown. These are produced by fitting the code output to the normalised experimental spectra using the vapour temperature as a fitting parameter.

The sub-plots in the figure \ref{results} show the deviations of the spectra taken at a temperature of 25~$^{\circ}$C from their respective fits to theory. It is clear to see that there is excellent agreement between theory and experiment across both of the $D_{2}$ lines. The rms deviation is at the 0.5~\% level; a typical level of agreement for the range of temperatures used.

Following minimal adaptation, the code for the electric susceptibility of atomic rubidium used by Siddons \textit{et al.} has been shown to successfully predict the Doppler-broadened absorption spectra of the caesium $D_{2}$ lines. We have thoroughly tested the code which evidently models the quantum system to a good level of accuracy. To improve this further, the magnetic octupole interaction of the caesium $D_{2}$ lines could be incorporated into the model \cite{Gerginov}. However this effect has a very minor contribution to the absorption lineshape and we feel that the current level of accuracy is sufficient. To extend the range of temperatures in which the model applies, a mechanism for collisional broadening would have to be included such as that by Weller \textit{et al.} \cite{weller}.

\subsection{The Off-Resonance Faraday Effect}
\begin{figure}[ht]
\centering
\includegraphics*[width=\linewidth]{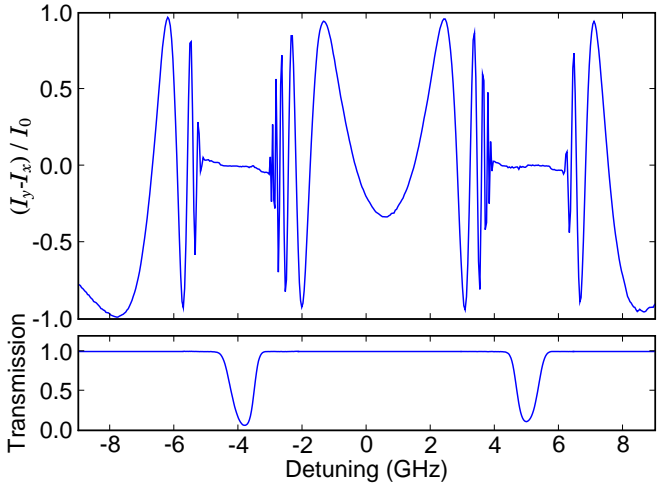}
\caption{An example Faraday signal across the entirety of the $D_{2}$ lines at a temperature of 71~$^{\circ}$C and a central magnetic field of 174$\pm$1~G. For comparison a transmission spectrum, across the same region, at a temperature of 19~$^{\circ}$C is shown in the sub-figure.}
\label{faradaysignal}
\end{figure}

A composite Faraday signal over the $D_{2}$ lines is shown in figure \ref{faradaysignal}. $I_{x}$ and $I_{y}$ are the intensities of the two orthogonal linear polarisations analysed by the PBS in figure \ref{setup}, $I_{0}$ is the intensity of the incident probe beam. For reference the absorption spectrum  across the $D_{2}$ lines at 19~$^{\circ}$C is shown in the sub-plot. These plots reveal several interesting features of the Faraday effect. Firstly, it is worth noting that the Faraday signal is very sensitive to detuning, even in the regions where close to 100~\% transmission is observed. This in itself makes the Faraday effect extremely useful since signals are produced far off-resonance without significant loss of light. Secondly, unlike rubidium, the large ground state hyperfine splitting in caesium leads to significant signal oscillation in the region between the two lines. This region should be ultra-sensitive to the distribution of atoms within the ground states. Consequently this frequency domain is a candidate for an all-optical switch \cite{dawes, siddons4}. Finally, one notices that the region of zero signal about the linecentres of the $D_{2}$ lines does not necessarily correspond to 100~\% absorption. A zero signal may be obtained with this optical set up when either the transmission of the light driving the $\sigma^{+}$ or $\sigma^{-}$ transition is zero. In both cases purely circularly polarised light emerges from the cell and will result in zero differencing signal.

\begin{figure}[ht]
\centering
\includegraphics*[width=\linewidth]{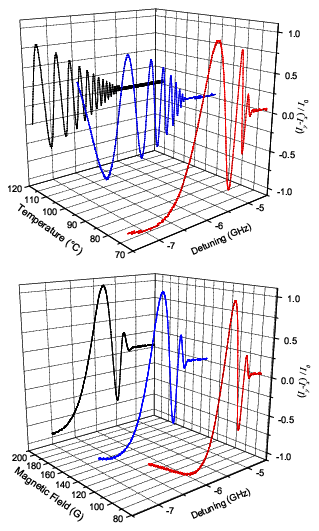}
\caption{Top: Evolution of Faraday signals with a constant magnetic field of 117~G with temperatures of 72~$^{\circ}$C (red), 92~$^{\circ}$C (blue), and 116~$^{\circ}$C (black). Bottom: Evolution of signals with a constant temperature of 61~$^{\circ}$C and magnetic fields of 89~G (red), 145~G (blue), and 210~G (black).}
\label{waterfall}
\end{figure}

Faraday signals at a variety of different temperatures and axial magnetic fields have been recorded. The temperature dependence of the Faraday effect manifests itself in the relationship between the electric susceptibility and the atomic number density. Magnetic field dependence is due to the splitting of the hyperfine sub-levels via the Zeeman effect. The waterfall plots of figure \ref{waterfall} demonstrate the evolution of the signals with either of these variables. We can immediately see that the vapour temperature has by far the greater impact on the signals; not only do the number of signal oscillations increase, but they also are shifted further from resonance. This sensitivity is due to the strong temperature dependence of the number density and, to a lesser extent, the width of the lineshape. Increasing the magnetic field simply shifts the edges of the lineshapes, corresponding to the translations of the signals seen in figure \ref{waterfall}.

\subsection{Modelling Far Off-Resonance Faraday Rotation}
The Zeeman effect can be included within the model described in section two by diagonalisation of the full coupling Hamiltonian \cite{Abel, Walther} in order to describe the Faraday signals. However, it will be shown in this paper that a simple model is sufficient to accurately predict the extent of Faraday rotation in a far off-resonance regime. When greatly detuned from an atomic resonance, dispersion increasingly dominates over absorption. It has been shown that, in this off-resonant region, the real part of the atomic lineshape factor may be approximated by \begin{equation} s_{r}\left(\Delta\right)\approx-1/\Delta,\label{approxdisp}\end{equation} where $\Delta$ is now the detuning from linecentre \cite{siddons}. Combining this with equation \ref{susceptibility}, the far-detuned real susceptibility becomes \begin{equation}\chi_{r}\left(\Delta\right)=-c_{j}^{2}\frac{d^{2}N}{\hbar\epsilon_{0}\Delta}.\end{equation} Since vapour cell number densities are generally very low ($\chi_{r}$ is typically much less than one), we can approximate the difference between the $\sigma^{+}$ and $\sigma^{-}$ refractive indices as \begin{equation} \eta^{+}-\eta^{-}=\sqrt{\chi_{r}^{+}+1}-\sqrt{\chi_{r}^{-}+1}\approx\frac{\chi_{r}^{+}-\chi_{r}^{-}}{2}.\label{refractivedifference}\end{equation} Again if we limit ourselves to a regime far from the atomic resonances, we may treat all of the constituent hyperfine transitions, in the single Doppler-broadened line, as a single transition of strength $C^{2}$. Furthermore we assume that the Zeeman effect translates into a shift of the refractive index profiles, $\eta^{+}$ and $\eta^{-}$, by a frequency $b$ in opposite directions. In this case equation \ref{refractivedifference} becomes \begin{equation}\eta^{+}-\eta^{-}=-C^{2}\frac{d^{2}N}{2\hbar\epsilon_{0}}\left[\frac{1}{\Delta+b}-\frac{1}{\Delta-b}\right].\end{equation} In the limit $\left|\Delta\right|>>b$; \begin{equation}\eta^{+}-\eta^{-}\approx C^{2}\frac{d^{2}Nb}{\hbar\epsilon_{0}\Delta^{2}}.\end{equation} Combining this with equation \ref{angle} the far off-resonant Faraday rotation angle is given by \begin{equation} \theta=C^{2}\frac{\pi Ld^{2}Nb}{\lambda\hbar\epsilon_{0}\Delta^{2}}.\end{equation} One of the many applications of the Faraday effect is the locking of lasers at large detunings from resonance \cite{Marchant}. Consider a set up with which locking points are generated at detunings where a rotation in the plane of polarisation of $n\pi$/2 is observed, where $n$ is an integer. By this simple model, with a constant magnetic field, the locking points are expected to occur at detunings \begin{equation}\Delta_{n}=\sqrt{\frac{AN\left(T\right)}{n}},\label{model}\end{equation} where $A$ is a fitting parameter characterising all the physical constants and approximations used, and $n$ is the order of the locking point. $N\left(T\right)$ is a the atomic number density as a function of temperature; a well-known function for all of the alkali metal atoms \cite{alcock}.

\begin{figure}[ht]
\centering
\includegraphics*[width=\linewidth]{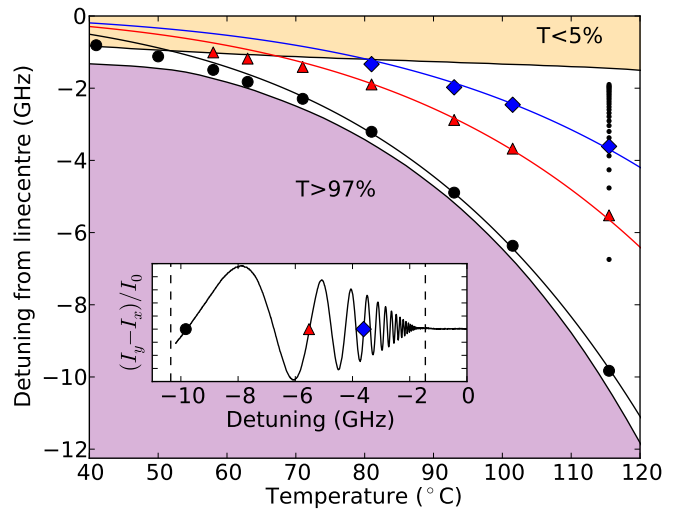}
\caption{The red-detuning from the \emph{F}=4 line of the 1$^{{\rm st}}$ (black circles), 3$^{{\rm rd}}$ (red triangles), and 7$^{{\rm th}}$ (blue diamonds) order zero crossings as a function of temperature for a central magnetic field of 117$\pm$1~G. The small black points indicate the locations of all discernible zero crossings on a 115~$^{\circ}$C signal. The blue curve is a fit to our model of the 1$^{{\rm st}}$ order crossings, the other curves are scalings of this fit. The shaded regions show the detunings at which there is less than 5~\% (orange) and more than 97~\% (purple) $\sigma^{+}$ transmission. The inset shows the signal recorded at 115~$^{\circ}$C and the locations of the various zero crossings.}
\label{contour}
\end{figure}

In order to test this simple analytic expression for far off-resonance Faraday rotation, we recorded the detunings of the signal zero crossings from the \emph{F}=4 linecentre as a function of temperature. The detunings of the first, third and seventh order zero crossings are plotted as a function of temperature in figure \ref{contour}, where the order is determined by the integer $n$ in equation \ref{model}. A fit of equation \ref{model} to the far-detuned first order zero crossings is also included. This curve has been scaled to the other data sets by dividing by the square root of the appropriate value of $n$. The inset shows a Faraday signal taken at 115~$^{\circ}$C with a central magnetic field of 117$\pm$1~G and  zero crossings corresponding to the data sets on the main figure are shown. We can see from the figure that at detunings greater than 2~GHz our model works excellently, with the data points deviating from the theory by no more than 1~\%. However, below 1.6~GHz there is very poor agreement with percentage deviation exceeding 30~\% in some cases. This behaviour is due to the approximations used within this theory that are only valid far from resonance. In reality the lineshape is much more complicated close to resonance than the simple expression used in equation \ref{approxdisp}.

Another phenomenon observed in figure \ref{contour} is that there appears to be a temperature dependent  threshold detuning,  below which there are no zero crossings. Our model fails to predict this since since it does not include the dichroism discussed earlier. The orange (light shading) region on the plot shows the region in which there is less than 5~\% $\sigma^{+}$ transmission, and in the purple (dark shading) region there is more than 97~\% transmission. These limits are also indicated by dashed lines on the inset. This defines a window in which which zero crossings should occur. We can see from the figure that it is indeed the case that no zero crossings occur outside this white region.

This simple model works exceptionally well at large detunings, which is the region it was designed to describe. This approximation can be used as a quick calibration tool for an inexpensive and simple method of locking lasers far from resonance; a necessity in optical dipole traps. In addition to the reuslts presented here, which has focussed on red-detuned light from the \emph{F}=4 line, we have verified that this model is successful at other magnetic field strengths for both lines. We have also used the model to successfuly predict locking points in the rubidium $D_{2}$ lines by fitting to the data presented by Marchant \textit{et al.} \cite{Marchant}.

\section{Conclusion}
In summary we have carefully investigated absorption and dispersion in hot caesium vapour. We have established the existence of a weak-probe regime in the caesium $D_{2}$ lines, in which the process of hyperfine pumping is negligible. In light of the identification of this regime, the Doppler-broadened $D_{2}$ absorption spectrum has been predicted for temperatures up to 50~$^{\circ}$C with a typical rms deviation of 0.5~\% from experimental data.

Whilst a code similar to that used in this paper may be used to predict Faraday rotation for applications such as laser locking, an analytical expression is desirable for purposes such as calibration. We have developed a simple model that may predict Faraday rotation at large detunings from resonance. This model has been demonstrated to be accurate to within 1~\% of experimental data, at detunings greater than 2~GHz from  resonance, for temperatures up to 115~$^{\circ}$C. The major advantage of this model is that only a single data set is required for calibration, and the other locking points may be extrapolated from a fit to the model. Consequently this model has great potential as a calibration technique for a simple and inexpensive method of laser locking far from resonance.

\begin{acknowledgements}
We are grateful to P. Siddons and L. Weller for assistance in adapting the code.
\end{acknowledgements}

\bibliographystyle{h-physrev}
\bibliography{refs}
\end{document}